\documentclass[a4paper,11pt]{article}

\usepackage{jinstpub}
\usepackage{lineno}
\usepackage{siunitx}

\newcommand{\nifs}{National Institute for Fusion Science, 322-6 Oroshicho, Toki, Gifu 509-5292, Japan}
\newcommand{\hokudai}{Division of Quantum Science and Engineering, Graduate School of Engineering, Hokkaido University, Kita 13, Nishi 8, Kita-ku, Sapporo, Hokkaido 060-8628, Japan}
\newcommand{\sokendai}{The Graduate University for Advanced Studies (SOKENDAI), 322-6 Oroshicho, Toki, Gifu 509-5292, Japan}
\newcommand{\mcdf}{swisdak13pop,zenitani15pop,zenitani22pop}

\proceeding{21$^{\text{st}}$ International Symposium on Laser-Aided Plasma Diagnostics\\
September 1--5 2025\\
Saillon, Switzerland}

\title{\boldmath Monte Carlo simulation method for incoherent Thomson scattering spectra from arbitrary electron distribution functions}

\author[a]{Kentaro Sakai}
\author[b]{Kentaro Tomita}
\author[a,c]{Takeo Hoshi}
\author[a,c]{Ryo Yasuhara}
\affiliation[a]{\nifs}
\affiliation[b]{\hokudai}
\affiliation[c]{\sokendai}

\emailAdd{sakai.kentaro@nifs.ac.jp}

\abstract{We developed a Monte Carlo simulation method to calculate incoherent Thomson scattering spectra in high temperature plasmas. The basic idea is to treat the entire scattering process as the superposition of individual photon-electron interactions. We introduce macro-particles, referred from particle-in-cell simulations, to reduce the computational cost, and obtain scattered spectra within a reasonable computational time. Since the velocity of the interacting electron is randomly sampled from an electron distribution function, the method can be applied to arbitrary electron distribution functions provided an appropriate sampling scheme is available. We present simulation results for relativistic Maxwellian and kappa distribution functions, and compare them with both analytical and numerical spectra for validation. The simulated spectra show good agreement with both analytical and numerical results, demonstrating that the Monte Carlo simulation method can reliably reproduce incoherent Thomson scattering spectra.}


\begin{document}
\maketitle
\flushbottom

\section{Introduction} \label{sec:intro}

Thomson scattering is a powerful diagnostics to measure local plasma parameters \cite{froula11}. In incoherent Thomson scattering, the projection of the electron distribution function along the measurement direction can be directly observed since this is Doppler spectroscopy of electrons \cite{van-lammeren92nf,tomita20jpd,shi22prl}. Although non-Maxwellian electron distribution functions can, in principle, be inferred from fine structures in the scattered spectrum, it is highly challenging to observe the plasma at low density and high temperature, where the collisional relaxation is weak and non-Maxwellian electron distribution function is expected to form, due to the limited signal-to-noise ratio of the measurement system. 
We are planning to observe non-Maxwellian electron distribution functions, involving anisotropy, deformations in bulk distribution functions, and non-thermal tails, using the Compact Helical Device (CHD), which is the upgraded device of the Compact Helical System \cite{matsuoka88iaea}. We have also shown the conceptual design of a high-wavelength-resolution Thomson scattering system capable of measuring non-Maxwellian with a fine signal-to-noise ratio \cite{sakai25ppcf}. As a next step, it is necessary to develop an analysis method to extract non-Maxwellian distribution functions. 

Because the scattered spectrum is not directly proportional to the electron distribution function except in the Maxwellian case \cite{tomita20jpd}, it is essential to understand the shape of scattered spectrum instead of the shape of electron distribution function itself. 
In order to implement non-Maxwellian electron distribution functions in the fitting process, the numerical method that can compute scattered spectrum in arbitrary distribution functions is required.
However, the analytical solutions of scattered spectrum are known for a few distribution functions \cite{palastro10pre,milder19pop,saito00angeo}, and arbitrary distribution functions are modeled as linear combinations of such models with many degrees of freedom \cite{sakai20pop,sakai23pop}. 
There are some other numerical methods to calculate Thomson scattering spectra using the randomly sampled electron trajectories \cite{pastor11nf,pastor12nf}, while it is computationally expensive. 
Motivated by the generation method to load various velocity distribution functions effectively in particle-in-cell simulations \cite{\mcdf}, 
we develop a Monte Carlo simulation method to calculate Thomson scattering spectra from arbitrary electron distribution functions, involving Maxwellian and all types of non-Maxwellian distribution functions. 
Section~\ref{sec:mc} describes the method of the Monte Carlo simulation, which is based on the idea of photon-electron scattering. 
Section~\ref{sec:sim} shows the Monte Carlo simulation results with a Maxwellian distribution function, and a kappa distribution function as an example of non-Maxwellian distribution functions. The simulated spectra show good agreement with theoretical spectra with high accuracy, indicating that this method can properly describe scattered spectra. 
In Sec.~\ref{sec:summary}, we provide a discussion and summary of our research.

\section{Monte Carlo simulation method} \label{sec:mc}

Thomson scattering is electromagnetic wave scattering by free electrons. The electrons oscillate in the incident electromagnetic waves, and the oscillating electrons act as electric dipoles, emitting dipole radiations. Since the energy of the incident wave is redistributed without being absorbed by the electron, the process is treated as an elastic electromagnetic wave scattering.  
Considering the fact that Thomson scattering is the classical limit of Compton scattering, it is also possible to treat the scattering process as a photon scattering by free electrons, instead of the conventional picture of electromagnetic wave scattering as a dipole radiation of an oscillating free electron in an incident electromagnetic field. The elementary process is a binary interaction between a single electron and a single photon, which can be treated as a stochastic process. By summing up the contributions of all photons and electrons considered within the system, we can obtain the scattered spectrum as a histogram of scattered photons.
The statistical treatment is particularly useful to simulate the experimental scattered spectra in high temperature plasmas, where the number of scattered photons in a single wavelength channel is small and statistical noise is significant. 

\begin{figure}
    \centering
    \includegraphics[clip,width=0.8\hsize]{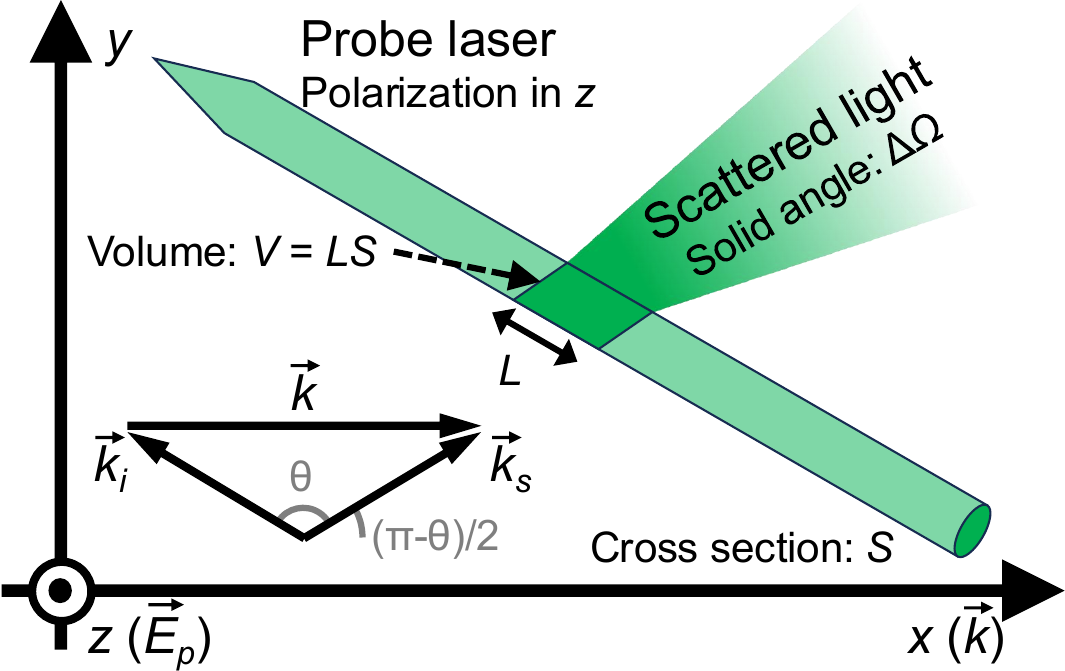}
    \caption{The scattering geometry within the scattering plane.}
    \label{fig:geometry}
\end{figure}

In this paper, we consider a scattering geometry in Fig.~\ref{fig:geometry}. The probe laser propagates within the $x-y$ plane and the angle between the $x$ axis and the incident wavevector $\vec{k}_i$ is $(\pi + \theta)/2$, where $\theta$ is the nominal scattering angle. The polarization of the probe laser $\vec{E}_p$ is aligned to the $z$ direction to maximize the Thomson differential cross section. The scattering occurs within the scattering volume $V=LS$, where $L$ is the integration length along the probe laser and $S$ is the cross section of the probe laser. The collection optics is located far away from the scattering volume, and the scattered wavevector $\vec{k}_s$ is tilted by $(\pi - \theta)/2$ from the $x$ axis. Therefore, the nominal scattering wavevector $\vec{k} = \vec{k}_s - \vec{k}_i$ is along the $x$ axis. Because the collection optics has a finite diameter, we observe the scattered light propagating within a finite solid angle $\Delta \Omega$. We define the unit vectors of incident wavevector, scattered wavevector, and laser electric field as $\vec{i}=\vec{k}_i/k_i$, $\vec{s} = \vec{k}_s/k_s$, and $\vec{p} = \vec{E}_p/E_p$, respectively.  

The scattering probability of Thomson scattering is governed by the Thomson differential cross section. The Thomson differential cross section with the relativistic aberration correction is given by \cite{williamson71jpp}
\begin{equation}
    \frac{d^2\sigma_T^{rel}}{d\omega_s d\Omega} = \frac{r_e^2}{\gamma^2} \frac{1-\vec{\beta}\cdot\vec{i}}{(1-\vec{\beta}\cdot \vec{s})^2} \left|1-\frac{ (1-\cos \theta) (\vec{\beta}\cdot \vec{p})^2}{ (1-\vec{\beta}\cdot \vec{i}) (1-\vec{\beta}\cdot \vec{s}) }\right|^2,
    \label{eq:cross_section}
\end{equation}
where $r_e$ is the classical electron radius, $\vec{\beta}=\vec{v}/c$ is the electron velocity normalized to the speed of light, and $\gamma = (1-\beta^2)^{-0.5}$ is the Lorentz factor. While the Thomson differential cross section is constant at $r_e^2$ in non-relativistic limit, that with relativistic aberration correction is a function of electron velocity. The differential cross section gives the scattering probability of a single electron in a unit frequency and solid angle. 
Note that the differential cross section does not consider a static magnetic field. If there is a strong magnetic field that significantly affect the electron trajectories, an additional correction due to the static magnetic field is required, which may result in a periodic wavelength spectrum \cite{nee69pof}. 
Using the differential cross section, the total number of scattered photons is given by
\begin{equation}
    N_{s} = \sum_{j=1}^{N_i} \sum_{k=1}^{N_e} \frac{\Delta \Omega }{S} \frac{d^2\sigma_T^{rel}}{d\omega_s d\Omega} \varepsilon (\lambda_s),
    \label{eq:photon_sim}
\end{equation}
where $N_i$ is the total number of incident photons, $N_e = n_e V$ is the number of electrons within the scattering volume, $n_e$ is the electron density, and $\varepsilon(\lambda_s)$ is the efficiency of the detection system as a function of scattered wavelength. This is set to unity when we calculate the shape of scattered spectrum, while use a realistic value comparing the results with the experimental ones. Note that the solid angle is used just to evaluate the amount of scattered photons. The effect of scattered wavevector change due to the finite solid angle is not included.
The electron velocity is different for individual electrons (subscript $k$). Because the differential cross section is a function of electron velocity, this term changes for each electron and is the origin of the relativistic aberration correction \cite{williamson71jpp}. 

Given the electron velocity, the Doppler-shifted wavelength of scattered photons ($\lambda_s$) can be calculated using 
\begin{equation}
    \lambda_s = \lambda_i\frac{1-\vec{\beta} \cdot \vec{s}}{1-\vec{\beta}\cdot \vec{i}}, 
    \label{eq:doppler}
\end{equation}
where $\lambda_i$ is the incident wavelength. The scattered wavelength is a function of both incident wavelength and electron velocity. 
Equation~\eqref{eq:doppler} is derived under the assumption of underdense plasmas, where the incident wave frequency is much larger than the electron plasmas frequency. Since the Doppler-shift is proportional to the incident wavelength, the scattered spectrum is also a function of the incident photon spectrum. In most cases of Thomson scattering system using a pulse duration with more than nanoseconds, the incident photon spectrum is approximated to monochromatic and the effect of incident photon spectrum is negligible, while using a short-pulse lasers of fs--ps duration, the effect of finite incident photon spectrum appears. This will be explained by the convolution of incident spectrum and Doppler shift in incoherent scattering using linear probe electromagnetic waves \cite{sakai25ctpp}. For fs--ps probe pulses, the convolution with the finite bandwidth of the incident spectrum should be included, which can be readily incorporated in this Monte Carlo framework. 

The basic strategy to calculate the Thomson scattering spectrum is to solve Eq.~\eqref{eq:photon_sim} with a given electron distribution function $f_e(\vec{v})$. However, both $N_i$ and $N_e$ is much larger than unity in many Thomson scattering systems. Assuming that a \SI{1}{J} pulse laser is operated at \SI{532}{nm}, $N_i \sim 3\times 10^{18} (E_L/\SI{1}{J}) (\lambda_L/\SI{532}{nm})$, where $E_L$ is the laser energy. The electron number with a \SI{e18}{m^{-3}} density in a $1\times 1\times \SI{1}{cm^3}$ volume is $N_e \sim 10^{12} (n_e/\SI{e18}{m^{-3}}) (V/\SI{1}{cm^3})$. These are typical parameters of imaging Thomson scattering system using the second harmonics of Nd:YAG laser \cite{han13rsi}. The expected number of scattered photons in the interaction between a single photon and electron determined by Eq.~\eqref{eq:photon_sim} is $\sim 4\times 10^{-28} (\Delta\Omega/\SI{0.1}{sr}) (S/\SI{1}{cm^2})^{-1} [d^2\sigma_T^{rel}/(d\omega_s d\Omega)/r_e^2] (\varepsilon/\SI{5}{\%})$. This is much less than unity and the calculation efficiency is low. In order to enhance the calculation efficiency, we use macro-particles of photon and electron, which is usually used in particle-in-cell simulations to reduce the computational resources. Using incident macro-photons and macro-electrons, Eq.~\eqref{eq:photon_sim} can be rewritten as
\begin{equation}
    N_{s} = \sum_{j=1}^{N_i/w_i} \sum_{k=1}^{N_e/w_e} w_i w_e \frac{\Delta \Omega }{S} \frac{d^2\sigma_T^{rel}}{d\omega_s d\Omega} \varepsilon(\lambda_s),
    \label{eq:photon_sim_memp}
\end{equation}
where $w_i$ and $w_e$ are the weight of a single incident macro-photon and a single macro-electron, respectively.
With a monochromatic probe laser, the incident laser wavelength is constant for all incident macro-photons and there is no need to treat photons stochastically. Putting $w_i = N_i$ into Eq.~\eqref{eq:photon_sim_memp}, we obtain 
\begin{equation}
    N_{s} = \sum_{k=1}^{N_e/w_e} N_i w_e \frac{\Delta \Omega }{S} \frac{d^2\sigma_T^{rel}}{d\omega_s d\Omega} \varepsilon(\lambda_s).
    \label{eq:photon_sim_me}
\end{equation}
The weight of macro-electron should be selected so that the summand in Eq.~\eqref{eq:photon_sim_me}, which is equivalent to the scattering probability, does not exceed unity. If it exceeds unity, the total number of scattered photons decreases as the weight becomes large. In order to conserve the total number of scattered photons, the weight is adjusted dynamically; if the summand is greater than unity, the macro-electron is split into two particles, each with half the weight. 
Using the scattering probability, the expected value of scattered photon spectrum can be written as 
\begin{equation}
    N_{s} (\lambda_s) = N_i w_e \frac{\Delta \Omega }{S} \varepsilon(\lambda_s) \sum_{k=1}^{N_e/w_e}  \frac{d^2\sigma_T^{rel}}{d\omega_s d\Omega}  f_e(\vec{v}_k) \delta \left(\lambda_s - \frac{1-\vec{\beta}_k\cdot \vec{s}}{1-\vec{\beta}_k\cdot \vec{i}} \right),
    \label{eq:photon_spec}
\end{equation}
where $\delta(x)$ is the delta function. 
Because $N_e$ is finite in the Monte Carlo simulation, Eq.~\eqref{eq:photon_spec} can be a discrete function. In the experiments, the measured spectrum is integrated over a finite wavelength interval. Thus, the scattered photon spectrum of the Monte Carlo simulation can be written as
\begin{equation}
    N_{s}^{MC} (\lambda_{s}) = N_i w_e \frac{\Delta \Omega }{S} \int_{\lambda_s-\Delta \lambda/2}^{\lambda_s+\Delta \lambda/2} \varepsilon(\lambda) \sum_{k=1}^{N_e/w_e}  \frac{d^2\sigma_T^{rel}}{d\omega_s d\Omega}  f_e(\vec{v}_k) \delta \left(\lambda - \frac{1-\vec{\beta}_k\cdot \vec{s}}{1-\vec{\beta}_k\cdot \vec{i}} \right) d\lambda,
    \label{eq:photon_spec_mc}
\end{equation}
where $\Delta \lambda$ is the wavelength integration interval. This is equivalent to creating a histogram of scattered photons using the bin consistent with the spectrometer setup. 

When $N_e\to \infty$, the sum in Eq.~\eqref{eq:photon_spec} is replaced by an integral and is proportional to
\begin{equation}
    N_{s}^{int} (\lambda_s) \propto \varepsilon(\lambda_s) \int_{|v|<c} \frac{d^2\sigma_T^{rel}}{d\omega_s d\Omega}  f_e(\vec{v}) \delta \left(\lambda_s - \frac{1-\vec{\beta}\cdot \vec{s}}{1-\vec{\beta}\cdot \vec{i}} \right) dv^3.
    \label{eq:photon_spec_int}
\end{equation}
Assuming $\varepsilon (\lambda_s) = 1$, Eq.~\eqref{eq:photon_spec_int} is equivalent to the original equation for the well-known scattered spectrum in high temperature plasmas \cite{matoba79jjap,selden80pla}.

\begin{figure}
    \centering
    \includegraphics[clip,width=\hsize]{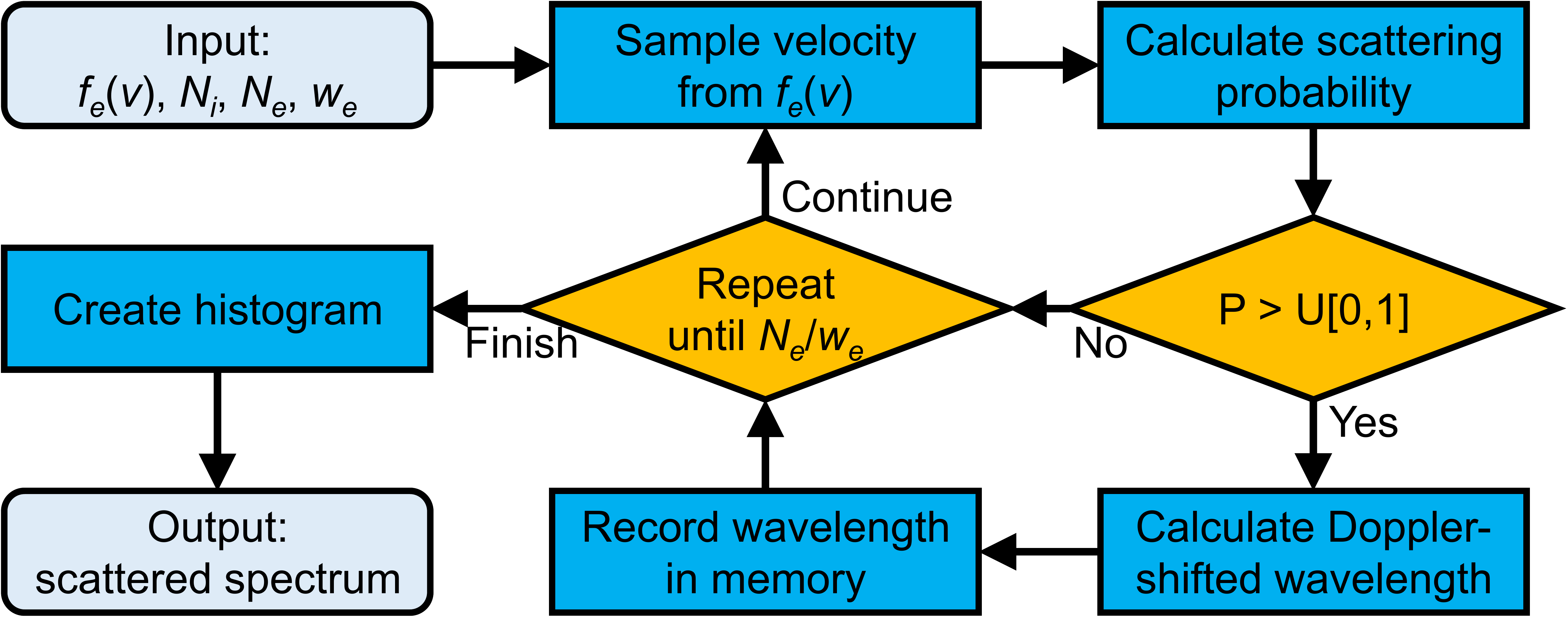}
    \caption{The flow chart of Monte Carlo simulation.}
    \label{fig:flowchart}
\end{figure}

Figure~\ref{fig:flowchart} shows the flow chart of the Monte Carlo simulation based on Eq.~\eqref{eq:photon_spec_mc}. As input parameters, we choose the electron distribution function, incident photon number, number of electron within the scattering volume, and the weight of a macro-electron. Because the differential cross section is a function of velocity, we sample a velocity from the electron distribution function using particle loading methods in particle-in-cell simulations \cite{\mcdf}. The sampled velocity is used to calculate the scattering probability and this is compared with a uniform random number between 0--1. If the probability is greater than the random number, we record as a scattered photon, otherwise not. This individual photon scattering process is repeated until $N_e/w_e$ and we output the histogram of scattered photons as a scattered spectrum. Since the spectrum is a histogram, the statistical noise of each channel is evaluated by the square root of the count. If $N_e\to \infty$, the statistical noise becomes negligible with respect to the signal level and the shape of scattered spectrum is properly described in the model.

Note that this method is invalid for collective scattering, i.e., the scattering $\alpha = 1/(k \lambda_D) \gtrsim 1$, where $\lambda_D$ is the Debye length, because the correlation of individual electrons are not taken into account in the stochastic process of scattering.

\section{Simulated spectra}\label{sec:sim}

\subsection{Maxwellian distribution functions} \label{sec:maxwell}

\begin{figure}
    \centering
    \includegraphics[clip,width=\hsize]{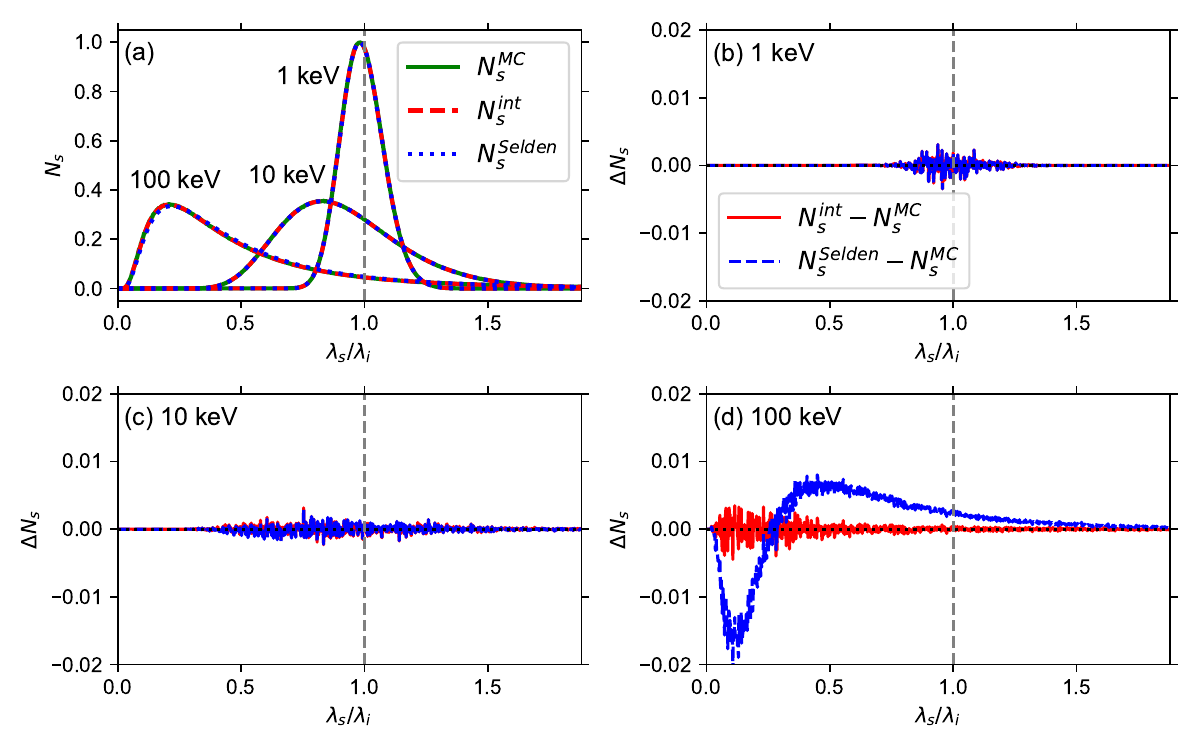}
    \caption{(a) Scattered photon spectra from relativistic Maxwellian distribution function obtained by Monte Carlo simulation, numerical calculation of Eq.~\eqref{eq:photon_spec_int}, and approximate analytical model with the scattering angle of \SI{163}{\degree}. (b)--(d) Difference between numerical result and approximate model and Monte Carlo simulation at $T_e =1$, 10, and \SI{100}{keV}. The legends in (b)--(d) are the same.}
    \label{fig:mc-selden}
\end{figure}

In order to verify the simulation model presented in Sec.~\ref{sec:mc}, we carried out a benchmarking of scattered spectrum using relativistic Maxwellian distribution functions. 
We compare the spectra calculated using Eqs.~\eqref{eq:photon_spec_mc} and \eqref{eq:photon_spec_int} with the approximate analytical spectrum shown in Ref.~\cite{selden80pla}. In the Monte Carlo simulation, we use a \SI{1}{J} laser operating at \SI{532}{nm}, which corresponds to \SI{1e18}{} incident photons. The scattering volume is $1\times1\times\SI{1}{cm^3}$, the solid angle of the collection optics is \SI{0.1}{sr}, and the nominal scattering angle is set to \SI{163}{\degree}. The wavelength channels of the spectrometer are arranged at \SI{1}{nm} intervals in the range of $0 < \lambda < \SI{1000}{nm}$. In order to calculate the shape of scattered spectrum, the detection efficiency is set to 1 for all wavelengths. We use the weight of macro-electron of \SI{1.2e8}{} and the electron density of \SI{1.2e21}{m^{-3}}, thus, the total number of macro-electrons is \SI{1e8}{}. Note that the electron density has no physical meaning. We use artificially high electron density to reduce statistical noise in the spectrum. Figure~\ref{fig:mc-selden}(a) shows the scattered spectra at different temperatures of $T_e = 1$, 10, and \SI{100}{keV}. The solid, dashed, and dotted curves correspond to the Monte Carlo simulation in Eq.~\eqref{eq:photon_spec_mc}, numerical solution of Eq.~\eqref{eq:photon_spec_int}, and approximate analytical spectrum derived by Selden \cite{selden80pla}, respectively. The vertical axis is normalized by the peak count of the \SI{1}{keV} spectrum. The dashed vertical line represents the incident wavelength. These three spectra are in good agreement with each other, indicating that the Monte Carlo simulation reproduces the scattered spectra with high accuracy. 

Figures~\ref{fig:mc-selden}(b)--\ref{fig:mc-selden}(d) show the difference of the Monte Carlo simulation result with the numerical solution of Eq.~\eqref{eq:photon_spec_int} (solid curve) and approximate analytical spectrum in Ref.~\cite{selden80pla} (dashed curve) at $T_e = 1$, 10, and \SI{100}{keV}, respectively. The solid curves in Figs.~\ref{fig:mc-selden}(b)--\ref{fig:mc-selden}(d) are close to 0, indicating that the Monte Carlo simulation agrees with Eq.~\eqref{eq:photon_spec_int} at the level of numerical error.
The noise level of Monte Carlo simulation is less than \SI{1}{\%} of the peak count in each temperature. While both the solid and the dashed curves are close to 0 at $T_e = 1$ and \SI{10}{keV}, that at $T_e=\SI{100}{keV}$ shows non-zero value, i.e., there is a discrepancy between the Monte Carlo simulation [and the numerical solution of Eq.~\eqref{eq:photon_spec_int}] and the approximate analytical spectrum. This indicates that the systematic error of Monte Carlo simulation is smaller than the approximate analytical model at \SI{100}{keV}. As mentioned in the literature \cite{selden80pla}, the approximate analytical spectrum is valid for $0.1\le T_e \le \SI{100}{keV}$ and shows $\approx \SI{1}{\%}$ difference from the numerical results at \SI{100}{keV}, which is consistent with the observed discrepancy. 

\begin{figure}
    \centering
    \includegraphics[clip,width=0.8\hsize]{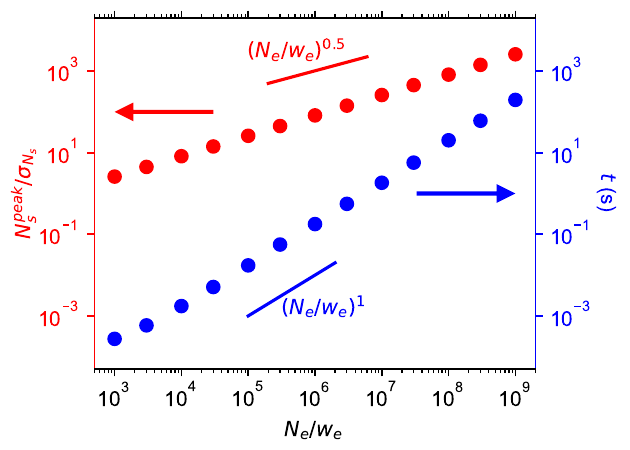}
    \caption{Signal-to-noise ratio and computational time as a function of number of macro-electrons at $T_e=\SI{10}{keV}$. The red and blue markers indicate the signal-to-noise ratio and the computational time corresponding to the left and right axes, respectively.}
    \label{fig:time}
\end{figure}

The signal-to-noise ratio of the Monte Carlo simulation increases as the number of macro-electrons increases. At the same time, the computational cost increases. We plot the signal-to-noise ratio and computational time of the Monte Carlo simulation as a function of number of macro-electrons at \SI{10}{keV} in Fig.~\ref{fig:time}. The red and blue markers indicate the signal-to-noise ratio and the computational time corresponding to the left and right axes, respectively. We define the signal-to-noise ratio as the ratio of the peak count to the standard deviation. The calculation is performed on a single thread of an Apple M4 Max processor. Because the individual scattering process of macro-electron and macro-photon is calculated in parallel, the computational time linearly scales to the number of macro-electrons. The signal-to-noise ratio is proportional to the square root of the number of macro-electrons. Note that the noise level, which is the inverse of the signal-to-noise ratio, is proportional to $(N_e/w_e)^{-0.5}$ and converges to zero with large number of macro-electrons.  
These suggest that the accuracy of scattered spectrum can be controlled by the hyperparameter of the simulation. 

\subsection{Non-Maxwellian distribution functions} \label{sec:nonmaxwell}

\begin{figure}
    \centering
    \includegraphics[clip,width=\hsize]{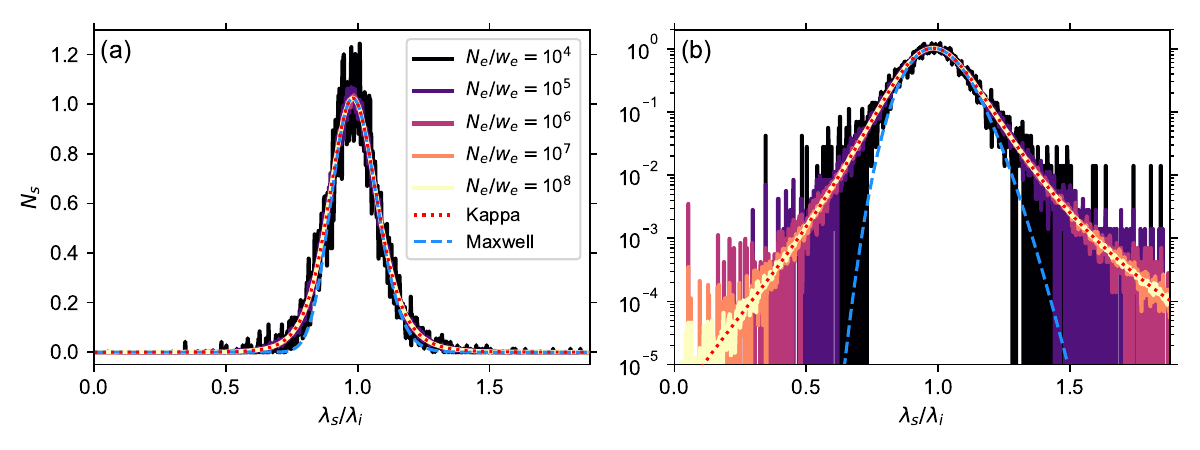}
    \caption{Scattered photon spectra from relativistic kappa distribution function with $T_e=\SI{1}{keV}$ and $\kappa = 3.5$ obtained by Monte Carlo simulation and numerical calculation of Eq.~\eqref{eq:photon_spec_int} in (a) linear and (b) logarithmic scales. The legends in (a) and (b) are the same.}
    \label{fig:kappa}
\end{figure}

We also carried out a benchmarking using non-Maxwellian distribution function. As an example, we use a relativistic kappa distribution function, which consists of thermal component and power-law tail \cite{pierrard10sp}, with $T_e=\SI{1}{keV}$ and $\kappa = 3.5$ in order to check the noise level at the edge of high-energy tail. Figure~\ref{fig:kappa} compares the Monte Carlo simulation and numerical calculation of Eq.~\eqref{eq:photon_spec_int}. The solid curves show the Monte Carlo simulation results, and dotted and dashed ones show the numerical spectra of relativistic kappa and Maxwellian distribution functions, respectively. The vertical axis is normalized by the peak count.  We change the total number of macro-electrons within the range of $10^4\le N_e/w_e \le 10^8$, which corresponds to the electron density range of $\SI{1.2e17}{}\le n_e \le \SI{1.2e21}{m^{-3}}$. As shown in Fig.~\ref{fig:kappa}(a), the simulated spectra of all cases quantitatively agrees with the numerical spectrum of the kappa distribution function. Even in the noisy case with $N_e/w_e=10^4$, there are no systematic errors in the simulated spectrum, in contrast to the numerical calculation of Eq.~\eqref{eq:photon_spec_int} with a rough interval of numerical integration. 
While it is difficult to distinguish kappa and Maxwellian distribution functions in the noisy case with $N_e/w_e=10^4$, one can find a deviation from the Maxwellian spectrum with $N_e/w_e\ge 10^5$, which is remarkable on a logarithmic scale in Fig.~\ref{fig:kappa}(b). 
When $N_e/w_e=10^8$, the simulated spectrum well agrees with the numerical spectrum even at the tail of the distribution function ($N_s \lesssim 10^{-4}$). This suggests that the Monte Carlo simulation model has an accuracy of more than four digits if the hyperparameters of the simulation are chosen appropriately. 

\section{Discussion and summary} \label{sec:summary}

As discussed in Sec.~\ref{sec:mc}, there are several assumptions in the model. For instance, the incident wave is assumed to be monochromatic, the scattered wavevector change due to the finite solid angle is neglected, and the instrumental function of the spectrometer is set to the delta function. These can be implemented as stochastic processes, e.g., the spectrum of incident wave can be included using Eq.~\eqref{eq:photon_sim_memp} with $w_i < N_i$. Further development is required to obtain realistic incoherent Thomson scattering spectra using the Monte Carlo simulation. 

In this paper, we use the Monte Carlo simulation as a calculation method to obtain incoherent Thomson scattering spectra without systematic errors. Another important aspect is to obtain scattered spectra with quantified statistical uncertainties. Because the amount of statistical uncertainties can be evaluated using the Monte Carlo simulation, this can be used to design a new Thomson scattering system. The setup of detection system can be implemented in the efficiency function of $\varepsilon(\lambda_s)$; this function includes, for example, transmittance and/or reflectance of optics, coupling efficiency of fibers, and quantum efficiency of detectors. This will enable quantitative comparison of systems, allowing the selection of the better system.
We consider the scattered photon spectrum because we are planning to use an image intensifier that counts photon number \cite{sakai25ppcf}. By multiplying the Thomson differential cross section in Eq.~\eqref{eq:cross_section} by $(1-\vec{\beta}_i)/(1-\vec{\beta}_s)$ \cite{williamson71jpp}, the Monte Carlo simulation method is easily extended to the scattered power spectrum, which is normally used in Thomson scattering measurements using filtered polychromator arrays \cite{yamada10fst}. Thus, this method is also applicable to the conventional systems.

\begin{figure}
    \centering
    \includegraphics[clip,width=\hsize]{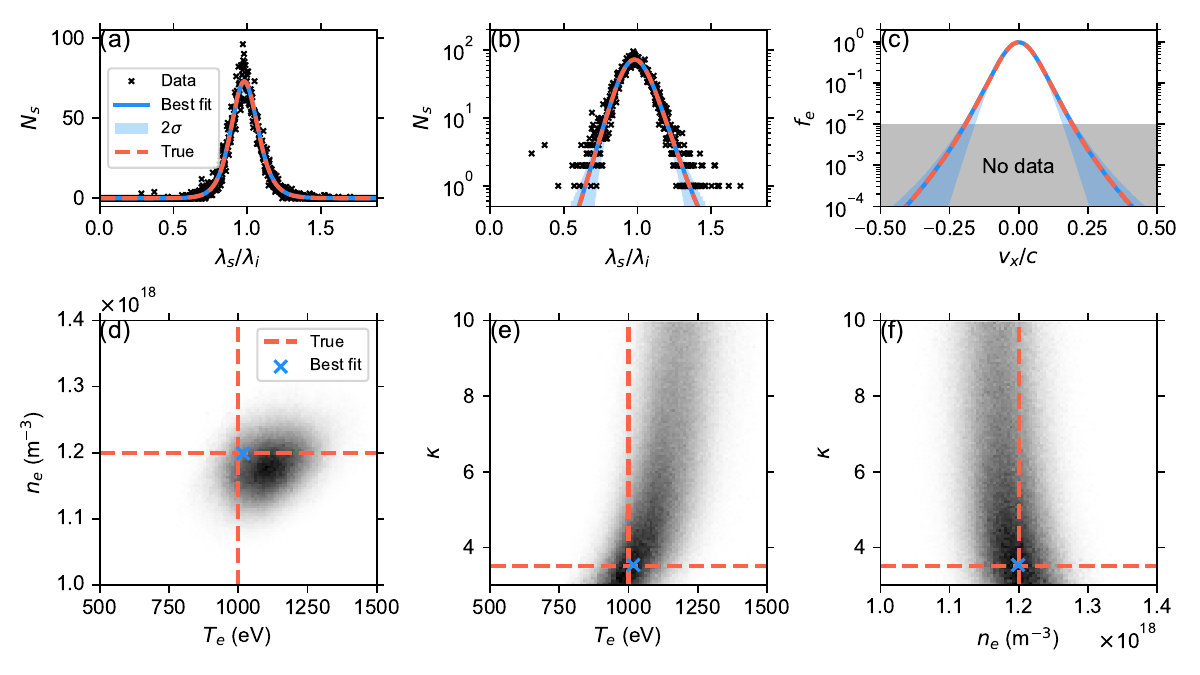}
    \caption{Bayesian inference assuming kappa distribution function. Scattered photon spectrum in (a) linear and (b) logarithmic scales. (c) Estimated electron distribution function. The legends in (a)--(c) are the same. (d)--(f) Two-dimensional projections of posterior probability density function. The color shows the probability density function. The legends in (d)--(f) are the same.}
    \label{fig:bayes}
\end{figure}

The estimated parameters are electron temperature and density in Maxwellian distribution function. Since the shape of scattered spectrum solely depends on the electron temperature, it is straightforward to evaluate both electron temperature and density; the temperature is estimated by one-dimensional exploration of the shape, and the electron density is estimated from the intensity. However, both shape and intensity are functions of two or more parameters in most non-Maxwellian models, which require nonlinear fitting algorithms. 
We are considering that the Monte Carlo simulation can be combined with the Bayesian inference to obtain the most plausible electron distribution function \cite{sakai25ppcf,motoyama22cpc}. While the scattered spectrum in Maxwellian distribution function is easily calculated using the approximate analytical model \cite{selden80pla}, there are only a few analytical models available for non-Maxwellian distribution functions \cite{palastro10pre,milder19pop,saito00angeo}. Because the computational cost of Monte Carlo simulation is not so high, the forward model of Thomson scattering measurement in the inverse problem analysis can be replaced by the Monte Carlo simulation. 
Figure~\ref{fig:bayes} shows an example of Bayesian inference of kappa distribution function. We generate an inference spectrum under the same condition as the $N_e/w_e=10^{4}$ case in Fig.~\ref{fig:kappa} but using a different random seed. 
We use scattered spectra calculated by the Monte Carlo simulation as the forward model of Thomson scattering spectra, and we minimize the the sum of squared residual between the inference spectrum and the simulated spectrum by changing the electron temperature, electron density, and spectral index ($\kappa$). 
The minimized sum of squared residual and the posterior probability density function of the parameters are obtained using the population annealing Monte Carlo sampling method \cite{hukushima03aipp} implemented in ODAT-SE (formally known as 2DMAT) \cite{motoyama22cpc}. The uniform prior probability density function is assumed to be uniform in the sampling domain of $500\le T_e\le 1500$~eV, $\SI{1e18}{}\le n_e\le\SI{1.4e18}{m^{-3}}$, and $3.5\le \kappa \le 10$.
As shown in Figs.~\ref{fig:bayes}(a) and \ref{fig:bayes}(b), the best-fit spectrum well agrees with the true spectrum in both linear and logarithmic scales. The $2\sigma$ confidence interval is increasing at the tail of the spectrum at $\lambda_s/\lambda_i\lesssim 0.7$ and $\lambda_s/\lambda_i\gtrsim 1.3$. The peak count is $\sim 100$~photons, i.e., the dynamic range is $\sim 10^2$. Figure~\ref{fig:bayes}(c) shows the estimated electron distribution function. The vertical axis is normalized by the peak count of the distribution function. The $2\sigma$ confidence interval agrees with the input distribution function at $f_e \gtrsim 10^2$, while its width increases at $f_e \lesssim 10^2$ because the dynamic range of the spectrum is $\sim 10^2$ and there is no effective data at the range. 
The posterior probability density function obtained from Bayesian inference is a probability density function in a three-dimensional space spanned by temperature, density, and spectral index. Figures~\ref{fig:bayes}(d)--\ref{fig:bayes}(f) show the probability density functions integrated over the spectral index, density, and temperature, respectively.
The projected posterior probability density functions in Figs.~\ref{fig:bayes}(d)--\ref{fig:bayes}(f) distribute around the true values indicated by the dashed lines. 
While the best-fit parameters are close to the true values, the peak of the probability density function in Fig.~\ref{fig:bayes}(d) is slightly shifted.
When Figs.~\ref{fig:bayes}(e) and \ref{fig:bayes}(f) are integrated over the spectral index, regions that do not correspond to the best-fit result appear to have higher probability density. However, in the full three-dimensional probability density function, the vicinity of the best-fit result exhibits the largest values. 
In Fig.~\ref{fig:bayes}(d), the discrepancy of the best-fit result and the peak in the probability density function arises from the limitations of the two-dimensional visualization. 
Because of the limited dynamic range, the power-law tail of the distribution function is less sensitive. With the larger $\kappa$, the high-energy components decrease and low-energy components increase, which leads to a systematic tendency to overestimate the temperature and underestimate the density. These demonstrate the usefulness of the Monte Carlo simulation as a forward model of inverse problem analysis. 

In the case of the kappa distribution function, the solutions are almost uniquely determined because similar distribution functions cannot be represented by different parameter sets. However, for more complex distribution functions, such as those expressed as the sum of multiple Maxwellian components \cite{sakai20pop,sakai23pop}, it is possible to represent the same distribution function using multiple distinct parameter sets. In such cases, the posterior probability density function is expected to exhibit a complex structure, potentially involving local maxima. 
By directly observing the posterior probability density function through Bayesian inference, multiple candidate velocity distribution functions can be identified. We expect that, by applying constraints derived from plasma physics knowledge to these candidate solutions, the most physically plausible distribution function can be reconstructed. Furthermore, the observed spectral width, skewness, and kurtosis of the spectrum can serve as useful constraints when selecting a non-Maxwellian velocity distribution model. How to incorporate these quantities into the data analysis remains an open issue for future work.
In order to further use this method in inverse problem analysis, it is also required to develop random sampling methods of various electron distribution functions \cite{\mcdf}. 

The method developed by Pastor et al. \cite{pastor11nf,pastor12nf} is more fundamentally grounded in first principle than that presented in this study, and it is expected to yield more accurate scattered spectra if the statistical noise originating from Monte Carlo simulations is negligible. Pastor's method directly solves the equation of motion for charged particles, allowing an accurate description of parameter regimes in which the laser intensity is high and the magnetic field contribution to the Lorentz force within the electromagnetic wave becomes non-negligible, i.e., regimes where electron trajectories exhibit so-called eight-figure motion rather than simple harmonic oscillation \cite{takabe20}. However, in such regimes, various nonlinearities arise due to laser-plasma interactions and laser energy is absorbed by the plasma to form nonlinear perturbations \cite{bierwage24srep}, making plasma diagnostics highly challenging \cite{sakai25ctpp}.
In contrast, the method presented in this paper employs a low-intensity laser and neglects effects from magnetic fields. This assumption is valid for most laser systems, except for those operating at extremely high intensities, such as Ti:Sapphire lasers, and is therefore considered reasonable. 
In terms of computational costs, Pastor's method requires 13h on a single CPU to simulate $10^{5\text{--}6}$ particles \cite{pastor11nf}, whereas the method presented in this study performs the same calculation in $\sim 0.1$~s (see Fig.~\ref{fig:time}). Even accounting for differences in computer performance, this represents a substantial reduction in computational costs. Such efficiency enables a dramatic suppression of Monte Carlo noise and allows for the calculation of low-noise scattered spectra in a fraction of the time. 
Nevertheless, for high-intensity laser applications, Pastor’s method or further development of more sophisticated techniques will be necessary.

In summary, we developed a Monte Carlo simulation method to calculate incoherent Thomson scattering spectra from arbitrary electron distribution functions. We consider the simulation model using the concept of photon-electron scattering, and the model is validated by comparing with the numerical spectra and approximate analytical model in Maxwellian and non-Maxwellian electron distribution functions, demonstrating the capability to compute incoherent Thomson scattering spectra in various non-Maxwellian distribution functions. Unlike purely analytical or deterministic numerical approaches, the Monte Carlo method not only avoids systematic errors but also provides a direct estimate of statistical uncertainties, which is particularly valuable for system design. This method can also be applied to the inverse problem analysis of real incoherent Thomson scattering spectra.

\acknowledgments

The computation in this work has been done using the facilities of the Supercomputer Center, the Institute for Solid State Physics, the University of Tokyo (2023-Ca-0122, 2024-Ca-0120, and 2025-Ca-0130).
This work is supported by JSPS KAKENHI (Grant Numbers 24K17029 and 23K25859), 
and by the NINS program of Promoting Research by Networking among Institutions (Grant Numbers 01422301 and 01412302). 
K.S. and T.H. are supported by the JST Moonshot R\&D Program (Grant Number JPMJMS24A3) for the data analysis using ODAT-SE.

\bibliographystyle{JHEP}
\bibliography{ref.bib}

\end{document}